# Electrically switchable cylindrical Fresnel lens based on holographic polymer-dispersed liquid crystals using a Michelson interferometer

H. Jashnsaz, E. Mohajerani*, and N. Hosain Nataj

Laser and Plasma Research Institute, Shahid Beheshti University, Evin, Tehran, Iran

*Corresponding Author Email: e-mohajerani@sbu.ac.ir

**Abstract—** Fabricating an electrically switchable cylindrical Fresnel lens based on holographic polymer-dispersed liquid crystals (H-PDLC) using a Michelson interferometer is reported. Simplicity of the method and possibility of fabricating different focal length lenses in a single set up are among the advantages of the method. It is demonstrated that the Fresnel structured zone plate acts as a cylindrical lens and focuses light in one dimension. Its electro-optical switching properties are also studied and it is found that at an applied electric field of E=6 $V_{rms}/\mu m$ across the sample, focusing property of the sample eliminates with a response of about 1 ms in a reversible manner.

## I. INTRODUCTION

Fresnel lens is a collapsed version of conventional lens and focuses light through diffraction instead of refraction. The Fresnel zone plate as a focusing element is more compact, possible to fabricate at large aperture size and easier to fabricate as compared with a lens of conventional design, thus resulting in a thinner, larger, and flatter lens with low light absorption and reduced required material as compared to conventional bulky lenses. A cylindrical Fresnel Lens, also referred to as a linear Fresnel Lens, has a linear structure, which can be used to collimate a row of light sources such as LEDs and to focus light in one direction only, resulting in a line image, where a concentric Fresnel lens focuses light in two dimensions and results in a point image. Controllable Fresnel lens with variable focal length or switchable focusing behavior is a critical component which has many applications in photonics such as optical interconnections, optical communications, imaging systems, displays, and light gathering applications [1]–[4].

Many approaches have been developed to fabricate Fresnel lenses by different materials; among them the electrically controllable liquid crystal Fresnel lens [5]-[10] is most promising because of good electro-optical properties of liquid crystals due to its molecular orientation in an external electric field. By combining liquid crystals with polymers and employment holographic techniques, a new composite material called holographic polymer-dispersed liquid crystals (H-PDLCs) [11] has emerged and electrically switchable holographic diffractive gratings [12]-[15] for various applications, using photopolymerizable liquid crystals polymer composite, have been demonstrated. Recently we have reported holographic techniques to fabricate electrically switchable LC/polymer Fresnel lens using Fresnel pattern resulted from superposition of a plan wave and spherical wavefront [16], [17]. Also we proposed and fabricated all-optical switchable holographic Fresnel lens using these techniques [18]. In this work we fabricate and characterize an electrically switchable Cylindrical Fresnel lens with focusing property in only one direction using an astigmatism lens in one path of Michelson interferometer, to our acknowledge this is the first study to propose and fabricate an electrically switchable one dimensional

Fresnel lens. In addition, we will demonstrate the possibility of fabricating cylindrical Fresnel lenses with different focal lengths and also astigmatism Fresnel lenses with different astigmatisms taking advantage of this method.

## II. THEORY

The consequence of superposition of plane wave and spherical wavefront is a concentric bright and dark rings of Fresnel pattern and irradiating of an LC/polymer composite film with this pattern yields us a concentric Fresnel lens which focuses light in two dimensions and results in a point image, Fig. 1(a), as discussed completely in our previous works [16, 17]. If one interfere a plane wave with a cylindrical wavefront instead of spherical wavefront, the result of interference is cylindrical (one dimensional) Fresnel pattern instead of concentric (two dimensional) bright and dark rings, as shown in Fig. 1(b).

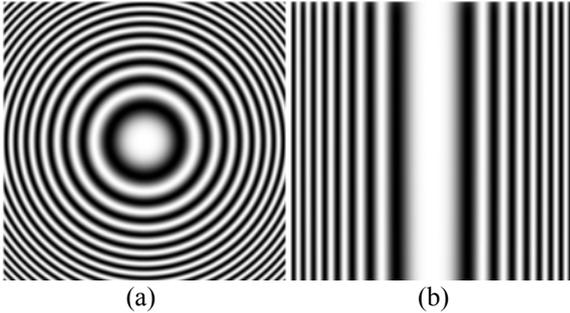

(a)                    (b)

**Fig**. 1 Fresnel zone plates; (a) concentric zone plate acts as a two dimensional Fresnel lens and focuses light to a point image and (b) cylindrical zone plate acts as a one dimensional Fresnel lens and focuses light to a line image.

When a photocurable homogeneous LC/prepolymer composite film is exposed to this pattern of light, photopolymerization process occurs in the bright regions faster than in the dark regions, and the monomer diffuses to the brighter regions while the LC molecules diffuses to the darker regions and form LC droplets. As a result of the Photo-polymerization induced phase separation (PIPS), concentration and size of liquid crystal droplets within the dark regions exceed those existing in the bright regions and a hardened polymer layers form in the bright regions, so, alternating layers of polymer and LC-droplet-rich regions form in the bright and dark regions, respectively. By choosing polymer in a manner to match its refractive index with one of the principal refractive index ($n_o$ or $n_e$) of the LC but not the other one, this composite film acts as a cylindrical Fresnel lens because of the refractive index mismatch between the adjacent zones in one dimension and it focuses light in one dimension only, resulting in a line image. When we apply an ac electric field across the film, the LC molecules are reoriented and the polymer network remains almost unchanged. Therefore, the relative phase difference between the zones $\Delta\varphi = 2\pi d \left(n_{LC} - n_{Polymer}\right)/\lambda$ alters. The focusing efficiency of the lens reaches its maximum value when the phase difference is equal to an odd multiple of $\pi$ while the film becomes isotropic for an even multiple of $\pi$.

## III. EXPERIMENTAL PROCEDURE

### A. *Sample Preparation*

We utilized materials formulation similar to that reported in our previous work [17]. The prepolymer liquid crystal-composite was made of commercially available materials. The materials used to fabricate the pre-polymer mixture were, trimethylolpropane triacrylate (TMPTA), as main monomer, N-vinylpyrrollidone (NVP), as crosslinking monomer, rose bengal (RB) as photoinitiator, N-phenylglycine (NPG), as co-initiator, and S-271 POE sorbitan monooleate as surfactant, all from Aldrich. The ratio of TMPTA/NVP/S-271/NPG/RB was 80/10/7/2/1 by weight. The refractive index of the cured polymer was 1.522 at 632.8 nm. After mixing prepolymer composites, the liquid crystal E7 with positive optical anisotropy from Merck and with the ordinary refractive index of $n_o$=1.521, and optical birefringence of $\Delta n$=0.225 was added to the mixture. The pre-polymer and LC were mechanically blended at 35/65 weight ratio, respectively, at 30 °C. Readymade homogeneous (HG) cells formed by two ITO coated glass plates separated by d≈9.9 μm thick spacer were filled with the photosensitive pre-polymer/LC homogenous mixture by means of the capillary flow.

## B. *Experimental Setup*

Figure 2 depicts the experimental setup of the Michelson interferometer we used to produce Fresnel pattern to cure the sample. Light from a cw argon ion laser with wavelength of $\lambda$=514 nm passes from a beam-expander then divided into two paths of the interferometer by a beam splitter (BS). Light propagates through path 2 as a planner wave, while it changes to an elliptical (approximately cylindrical) wavefront in path 1 by placing an astigmatism lens (with first focal length of F1≈100 cm) between BS and mirror 1. As a result of the superposition of these planner and elliptical waves, the elliptical Fresnel pattern forms in the plane perpendicular to the propagation axis in the output of the interferometer, and its central ellipse's wide (2a) and length (2b) alters by a displacement toward or away from the BS according to the $a=(2\lambda d_1)^{1/2}$ and $b=(2\lambda d_2)^{1/2}$ [18], where $\lambda$=514 nm, and focal lengths of astigmatism lens $L_3$ is responsible to $d_1$ and $d_2$. The variable *a* is one-half the length of the central ellipse's major axis; *b* is one-half the length of the minor axis. This gives a binary phase Fresnel lens with two different focal lengths $f_1 \equiv f(x)$ and $f_2 \equiv f(y)$ in direction of *x* and *y*, respectively and they are related to the inner most zone diameters *a* and *b* as $f_1 \equiv f(x)=a^2/\lambda$ and $f_2 \equiv f(y)=b^2/\lambda$, where $\lambda$ is the wavelength of the incident probe beam. Thus various Fresnel lenses with different focal lengths can be fabricated by this arrangement only by moving the sample toward or away from the BS. Also different astigmatism Fresnel lenses with different ratio of $f_1/f_2$ can be fabricated by using different astigmatism lenses (different F1/F2) at the place of L3. Polarizer P determines the polarization state of the curing beam. The power of laser was 120 mW over the sample, the exposure time was 4 minutes and the time of the final uniform UV curing was about 3 minutes. The experiment carried out on a vibration-isolated optical bench at room temperature. The central zone was dark, so the odd and even rings shown in the Fig. 2 became liquid crystal and polymer domain layers, respectively.

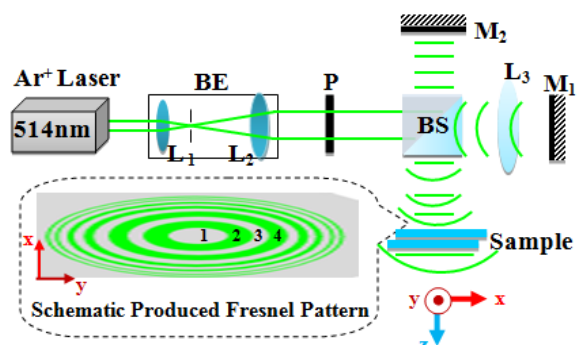

**Fig. 2** (Color online) Michelson interferometer; experimental arrangement for fabricating the holographic Fresnel lens; BE: beam expander; BS: beam splitter; L1, L2: lenses; L3: astigmatism lens (approximately cylindrical lens); M1 and M2: mirrors; P: polarizer.

Figure 3 depicts the experimental setup we use to study the performance and electro-optical properties of the fabricated lens. Expanded He-Ne laser beam with wavelength of 632.8 nm covers the area of the structured sample in a manner to experience Fresnel zone plate in x direction only, but not in y direction and in a rather good approximation it can be regarded such as Fig. 1(b) for incident probe beam, and it acts as a one dimensional Fresnel lens. To demonstrate its light focusing behavior a CCD Camera and a photodiode detector (PD) was set at focal length of the lens.

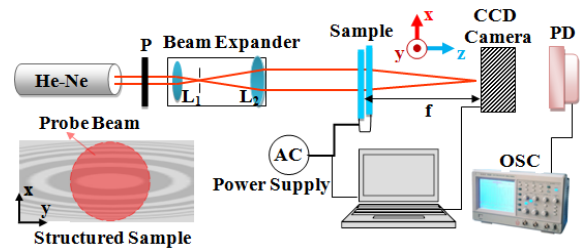

**Fig. 3** (Color online) Experimental arrangement for studying the light focusing behavior and electro-optical properties of the fabricated lens; L1 and L2: lenses; P: polarizer; PD: photodiode detector; schematic structure of patterned sample and its cross area with probe beam has shown in xy plane.

The data were analyzed by a computer and the output of the PD monitored on an oscilloscope (OSC). In order to study the electro-optical behavior of the sample, a 5 kHz ac electric field was applied across the sample. Polarizer P sets the polarization state of the probe beam in x direction which is the direction of the surface rubbing of the HG cell.

## C. Results and Discussion

Optical polarizing microscope images of the structured sample are shown in Fig. 4. The polymer zones form an isotropic network which cannot rotate the polarization of incident light, however, in the LC-droplet rich layers, nematic molecules form an anisotropic regions and show birefringence behavior. Therefore, the polarization of the light will be rotated after passing through LC layers. Thus under the crossed-polarizer optical microscope, the odd and even zones look bright and black which correspond to the anisotropic LC and isotropic polymer domain layers, respectively as shown in Fig. 4(a). The reverse is true if the microscope is under the parallel-polarizer condition as shown in Fig. 4(b). The central ellipse shaped zone is estimated to be about 0.5 mm width, 2a=0.5 mm.

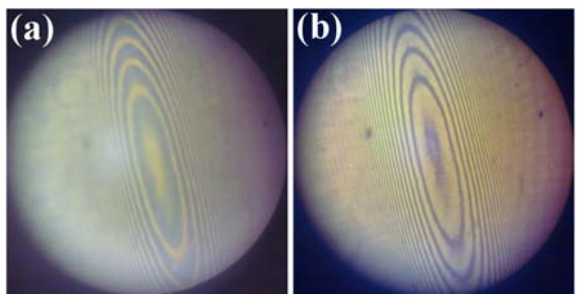

**Fig. 4** (Color online) polarizing microscope images of the patterned sample with (a) crossed polarizers, (b) parallel polarizers.

Figure 5 shows the focusing property of the fabricated lens using a CCD camera set at focal point. Incident probe laser beam profile has shown in Fig. 5(a) in the absence of sample. When the sample is present, focusing behavior occurs in one dimension, in x direction, as shown in Fig. 5(b), which has captured by a CCD camera set at about 8 cm from the sample at its focal point. This is due to different refractive index of the LC and polymer domain layers for incident probe light which acts as a one dimensional Fresnel zone plate because of refractive index mismatch between adjacent layers, so, it focuses light in x direction, resulting in a line image. These results indicate that the sample indeed behaves like a cylindrical lens.

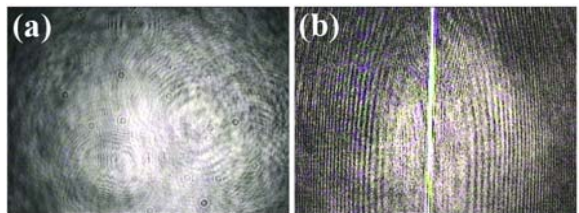

**Fig. 5** (Color online) the observed laser beam images (a) without LC sample, (a) with LC sample, using a CCD camera at focal point at λ=623.8 nm.

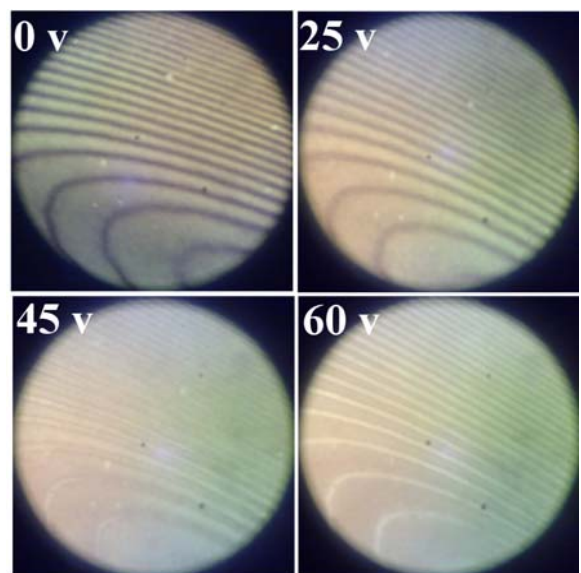

**Fig. 6** (Color online) microscope images of the patterned sample at various applied voltages, (Polarizer ∥ Analyzer).

Figure 6 shows the optical polarizing microscope images of the sample with parallel polarizers at various applied voltages. It look that by increasing applied electric field across the film, the dark layers in which the concentration of LCs is high gets brighter where there is not serious change in the brightness of the bright isotropic polymeric layers, and the intensity contrast between the dark and bright regions decreases and eventually at about 60 $V_{rms}$ applied voltage across the sample the initially dark layers completely gets bright and the sample becomes isotropic. The contrast between the dark and bright regions corresponds to the refractive index mismatch between adjacent LC rich and polymer layers.

Figure 7 shows Intensity profiles of the outgoing focused beam using a CCD camera set at focal point of the sample at different applied voltages. CCD camera captured picture without applied voltage is given inset the figure and profiles at various voltages all have yield along indicated horizontal dotted yellow line (analyzed in Origin Pro.).

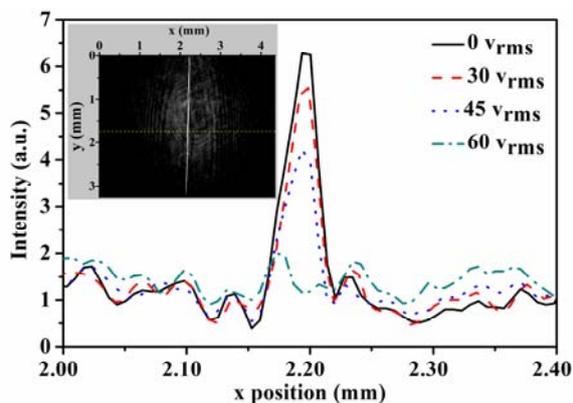

**Fig. 7** (Color online) intensity profiles of the outgoing focused (in one direction, x direction) beam by a CCD camera set at focal point of the sample at various applied voltages. CCD camera captured picture without applied voltage to sample is given inset the figure and profiles at various voltages all have yield along indicated horizontal dotted yellow line.

Figure 8 shows the diffraction efficiency at various applied electric field to the sample by a photodiode detector set at the focal point of the fabricated lens. By increasing the applied electric field across the sample, the focusing property of the sample decreases because of refractive index matching between LC and polymer layers, and eventually the diffraction efficiency eliminated at the 6 $V_{rms}/\mu m$, which corresponds to the switching voltage. This is the state in which the uniaxial rode-like LC molecules have reoriented in direction of the applied electric field (perpendicular to the cell surface), so transversely polarized light passing through LC layers experience the ordinary ($n_o$) refractive index of LC, which is equal to the refractive index of the polymer layers.

We also have measured the response times of the fabricated Fresnel lens by switching the applied electric field (60 $V_{rms}$ across the cell) on and off for rise and decay times, respectively. To perform this, a photodiode detector was set at the focal point and its output monitored on an oscilloscope, the results have been depicted below in the Fig. 9. The rise and decay response times both obtained about 1 ms.

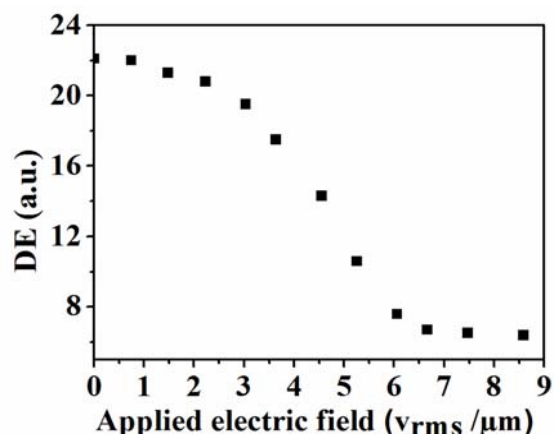

**Fig. 8** Diffraction efficiency at various applied electric field measured by a photodiode detector set at primary focal point of the sample.

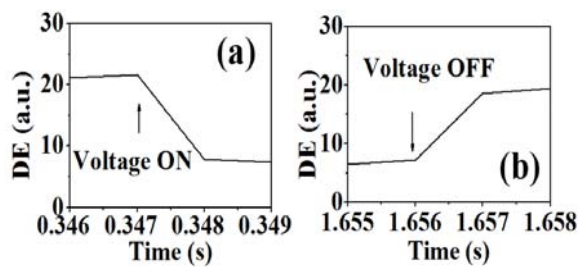

**Fig. 9** Electro-optical response time of the fabricated Fresnel lens monitored on an OSC using a photodiode detector at focal point (a) the rise time and (b) the decay time, by switching on and off an electric field of E=6 $V_{rms}/\mu m$ across the sample.

## IV. CONCLUSION

This study proposed and fabricated an electrically switchable Cylindrical Fresnel lens based on holographic polymer dispersed liquid crystals using a Michelson interferometer as a new holographic recording technique. One dimensional focusing property of the fabricated cylindrical Fresnel lens demonstrated and its electro-optical switching properties investigated. It is found that $E=6$ $V_{rms}/\mu m$ was switching voltage, and response time of 1 ms was obtained for both rise and decay times.